\documentclass[11pt,twoside]{article}

\usepackage{asp2006}
\usepackage{graphicx}

\begin{document}
\title{Galaxy Fomation}   %%% Fill in title
\author{Takashi Okamoto}   %%% Fill in author names
\affil{The Institute of Computational Cosmology, Department of Physics, Durham University, South Rodad, Durham, DH1 3LE, UK}    %%% Fill in author affiliations

\begin{abstract} %%% Abstract to run on from here.

I review the current status of theoretical studies of galaxy formation.
I outline the importance of the physics of baryonic component in galaxy 
formation by showing results obtained by using two major tools, 
semi-analytical approaches and cosmological simulations.  
In particular, I emphasis on roles of feedback in galaxy formation and 
discuss whether apparent conflictions between the standard theory of structure 
formation, the cold dark matter model, and observations can be solved by 
the feedback. 
I also discuss future prospects in numerical simulations of galaxy formation.

\end{abstract}

%%% MAIN BODY OF TEXT GOES HERE. CONSULT "INSTRUCTIONS FOR AUTHORS USING
%%% LATEX2E MARKUP", SECTIONS 2.3-2.6 FOR HELP WITH EQUATIONS, FIGURES,
%%% AND TABLES.

\section{Introduction}   %%% Top level section head (remove "%" symbol)

Understanding galaxy formation is a challenging problem whose solution will require a concerted approach combining observational, semi-analytical and numerical work.
Perhaps the most important progress in this area is the establishment of cold dark matter (CDM) model. The CDM model has steadily gained acceptance since it was first mooted in the early 1980s \citep{pee82, blu84, dav85}. 
This model has a tremendous predicting power and many of these predictions have turned out to be successful. 
Strong supports for this model come from the measurement of temperature anisotropies in the cosmic microwave background \citep[e.g.][]{smo92, spe07} and two large galaxy surveys: the two-degree Field Galaxy Redshift Survey \citep[2dFGRS][]{col01} and the Sloan Digital Sky Survey \citep[SDSS][]{yor00}, which have permitted the most accurate measurements to date of the power spectrum of galaxy clustering \citep[e.g.][]{per07, teg04, teg06, pop04, col05, pad07}. 
 This model specifies the values of the fundamental cosmological parameters and gives the initial condition for structure formation of the Universe such as galaxies.
We are now able to study galaxy formation by {\it ab initio} approaches based on the CDM model where the structure formation is characterized by the hierarchical clustering.

The second key advance that makes progress in understanding galaxy formation possible is the unveiling of the high redshift universe. 
Observations of galaxies over a range of redshifts allow us to compare their properties at different epochs in the history of the Universe. 
The unprecedented faint imaging of galaxies by the Hubble Deep Field \citep{wil96}, combined with the Lyman-break dropout technique to isolate high redshift galaxies \citep{ste96}, was essential in making possible the first determination of the cosmic star formation history \citep{mad96, ste99}. 
The detection of the emission from galaxies at sub-millimeter wavelengths \citep{sma97, bar98, hug98} offers the chance of uncovering up heavily dust obscured galaxies which are too faint to appear in optical surveys.

While it is the observations that has led the progress in this field, it is the theory that is needed to connect observed {\it snapshots} of evolving galaxy population. 
Our lack of understanding of the key physics underpinning galaxy formation, such as star formation and feedback, enforces us to introduce some phenomenological models.
I will illustrate the importance of such baryonic physics and how we are beginning to get to grips with it. 

\section{Formation of dark matter halos}

Initial tiny density fluctuations grow owing to gravitational instability and eventually form virialized objects, called dark matter halos. 
In a CDM universe, the process of perturbation growth is dissipationless. This means that the total kinetic energy of a system of dark matter retained, although energy can be converted from potential to kinetic. 
Hence, studying the growth of dark matter structure in a CDM universe is essentially straightforward to do by cosmological $N$-body simulations.  

\subsection{The abundance of dark matter halos}

The first attempt to calculate the abundance of gravitationally bound objects was mad by \citet{ps74} by assuming a Gaussian density field and smoothing the field on different scales. 
The mass function predicted by this simple calculation is agrees surprisingly well with the results obtained from $N$-body simulations \citep[e.g.][]{efs88, lc94}. 
\citet{smt01} presented a model in which they replaced the spherical collapse model with an ellipsoidal collapse. 
\citet{jen01} established the mass function of dark matter halos using a suite of $N$-body simulations and proposed a fitting formula that encapsulates the numerical results. This fitting formula was found to produce good fitting to down to 
$\sim 10^{10} M_{\odot}$ \citep{spr05}.

By utilizing the halo mass function, one can derive a mass-to-light ratio that guarantees a match between the theoretical prediction and the observed group luminosity function. 
A mass-to-light ratio that guarantees a match between the theoretical prediction and the observed group luminosity function is a strong function of halo mass \citep{yan03, eke04}. 
It is lowest for halos lf mass $\simeq 10^{12} h^{-1} M_{\odot}$ and rises by a factor of $\simeq 6$ to lower and higher mass. To understand this mass dependence of galaxy formation, we should consider baryonic physics.

\subsection{Density profiles of dark halos}

The increasing computer power and the advent of new simulation techniques allow to study the formation of individual dark matter halos in full cosmological context (i.e. from CDM initial conditions). 
Multi-mass simulation techniques, so-called {\it zoom simulations},\citep{nb91, kw93, fre96} resolve individual dark halos with $\sim 10^4$ particles \citep{nfw96, nfw97}, or even $\sim 10^6$ particles \citep{moo98, kly99a, oh99}.

NFW found that the density profiles of dark halos follow a universal law:
\begin{equation}
\rho(r) = \frac{\rho_0}{(r/r_s) (1 + r/r_s)^2}, 
  \label{nfw}
  \end{equation}
  where $r_s$ is a characteristic radius. 
  The profile is characterized by the inner slope $\alpha = d\ln \rho/d\ln r = -1$ for $r \ll r_s$ and the outer slope $\alpha = -3$ for $r \gg r_s$.
While many authors have confirmed that the NFW profile provides good fit, simulations with higher resolution form halos with inner slopes as steep as $\alpha = -1.5$ \citep{moo99b, fm01}.
  \citet{hay04} has claimed that there is no universal slope for the inner density profile and the inner slopes are steeper than NFW and shallower than Moore's, i.e. $-1.5 < \alpha < -1$ for $ r \ll r_s$. 

  The existence of such a universal density profile makes strong predictions on the shape of the rotation curve of galaxies that can be confronted against observations.
  Such comparisons have revealed that the inner slope of the density profiles of $\alpha < -1$, seems too steep and inconsistent with rotation curves studies of dark matter dominated galaxies, i.e. low surface brightness galaxies \citep{md98,fp94,moo99b,db02}.
  Some authors have argued that the rotation curves are consistent with CDM models if one considers the triaxiality of dark matter halos \citep{hn06, vel07, hns07}.
  However we should consider effects of baryonic component properly since the presence of a galaxy disk at the center of a dark matter halo makes dark halo significantly rounder than dark matter halos formed in dark matter only simulations \citep{bai05} and adiabatic contraction due to disk formation makes dark matter halos more concentrate as we shall see later.   
  The problem is not only at the very center of low surface bright galaxies but also seen at intermediate radii of any disk galaxies \citep{ns00a, mcg07}.
  This problem in the CDM density profile is called the {\it cuspy core problem}. 

\subsection{Substructure within dark matter halos}

Until the late 1990s, cosmological $N$-body simulations suggested that dark matter halos were smooth and featureless \citep{sum95, fre96}. 
Higher resolution simulations which typically contained more than $10^6$ particles in each dark matter halo showed that this phenomenon, called {\it over-merging}, was a numerical artifact \citep{ghi98, moo98, kly99a, oh99}. Once a halo enters within the virial radius of a more massive halo, it is referred to as a satellite halo or substructure within the larger halo ({\it sub-halo}).  
\citeauthor{oh99} (\citeyear{oh99}, \citeyear{oh00}) provided a method to construct merging histories of individual sub-halos in order 
to follow their formation and evolutions.   
\citet{moo99a} and \citet{kly99b} pointed out that the number of satellite halos in a Milky Way-sized halo is an order of magnitude larger than the number of the satellite galaxies of the Milky Way.
This problem is called the {\it satellite problem}.

\section{Semi-analytic models}    %%% Unnumbered top level section head (remove "%" symbol)

A powerful way of studying hierarchical galaxy formation is semi-analytic modelling of galaxy formation. 
\citet{wr78} proposed that galaxy formation was a two stage process, with dark halos forming in dissipationless, gravitational collapse, with galaxies forming inside these structures, following the radiative cooling of baryons. 
They also argued the importance of an additional process, feedback, for avoiding the production of more faint galaxies than are observed. 
\cite{wf91} produced a galaxy formation model that included many of today's models: CDM, radiative cooling of gas, star formation, feedback, and stellar populations. 
\citet{kwg93} and \citet{col94} presented the first models to track the formation and evolution of galaxies in the setting of evolving dark matter halos.
Below I briefly describe procedures that are commonly used in semi-analytic models.
A more comprehensive review on semi-analytic galaxy formation models can be found in \citet{bau06}.

\subsection{Basic ingredients}

\subsubsection{Dark halo merger trees}

Semi-analytic models require the formation history of each dark matter halo, commonly called the merger tree, in which a galaxy forms and evolves.
Merger trees can be constructed using a Monte-Carlo approach by sampling the distribution of progenitor masses predicted by the extended Press-Schechter theory \citep{lc94, sk99, col00}. 
More faithful way to construct merger trees is extracting from $N$-body simulations which have sufficiently frequent outputs. 
The first attempt to extract merger trees from $N$-body simulations was made by \citet{rou97}. 
\citet{kau99} studied galaxy distribution using merger trees generated by $N$-body simulations. 
\citet{on01} utilizes merger trees of sub-halos obtained in high-resolution cosmological $N$-body simulations of clusters of galaxies \citep{oh99, oh00} to investigate the morphology-density relation of cluster galaxies \citep{dre80}. 
This powerful method was firstly used for semi-analytic studies of cluster galaxies \citep{on01, on03, spr01} and has become a standard technique for semi-analytic models combined with $N$-body simulations \citep{cro06, bow06}.

\subsubsection{Gas cooling}

The basic model of how gas cools inside dark matter halos was set out in detail by \citet{wf91}. 
Semi-analytic models assume that when a dark matter halo forms, gas contained in the halo is shock heated to the virial temperature of the halo:
\begin{equation}
T_{\rm vir} = \frac{1}{2}\frac{\mu m_{\rm p}}{k} V_{\rm c},
  \end{equation}
  where $\mu$ is the mean molecular weight of the gas, $m_{\rm p}$ is the mass of a hydrogen atom, $k$ is Boltzmann's constant, and $V_{\rm c} = (G M_{\rm vir}/r_{\rm vir})^\frac{1}{2}$ is the circular velocity of the halo with mass $M_{\rm vir}$ and virial radius $r_{\rm vir}$.  
This gas with $T = T_{\rm vir}$ is called the {\it hot gas}. 
It is also assumed that the hot gas distributes parallel to the dark matter.
The relation between the circular velocity and halo mass is a function of redshift and cosmology. Gas can subsequently cool from the hot halo. 
The cooling rate of the gas is dependent on the temperature and metallicity of the gas. Hence a cooling time can be specified by dividing the thermal energy density of the gas by the cooling rate per unit volume:
\begin{equation}
t_{\rm cool}(r) = \left(\frac{3}{2}\frac{\rho_{\rm g} k T_{\rm vir}}{\mu m_{\rm p}}\right)/(\Lambda(\rho_{\rm g}, T_{\rm vir}, Z)),
  \label{COOLING}
\end{equation}
where $\rho_{\rm g}$ is the gas density which is a function of radius and the function $\Lambda$ is a cooling function. 
The cooling radius, $r_{\rm cool}$ is computed as the radius where the cooling time is equal to the time-step used in the merger tree. 
The gas within $r_{\rm cool}$ accretes to the center of the halo at rate set by the cooling time (or dynamical time if the dynamical time is longer than the cooling time) {\it keeping its angular momentum}. 
The cooled gas is assumed to have $\simeq 10^4$ K at which the atomic cooling becomes inefficient and this gas is called the {\it cold gas}. 

\subsubsection{Star formation}

The lack of a theory of star formation enforces and allows a simple estimate of the global rate of star formation in a model galaxy:
\begin{equation}
\dot{M}_* = \frac{M_{\rm cold}}{\tau_*},
\end{equation} 
where the star formation rate, $\dot{M}_*$, depends on the amount of cold gas available, $M_{\rm cold}$ and a star formation timescale, $\tau_*$. 
The star formation timescale is usually represented as 
\begin{equation}
\tau_* = \tau^0_* \left(\frac{V_{\rm c}}{V_*}\right)^{\alpha_*},
\end{equation}
where $\tau^0_*$ is either the dynamical time $\tau_{\rm dyn} = r_{\rm gal}/v_{\rm gal}$ within the galaxy, which is a function of redshift, or some constant and the dependence on the circular velocity is introduced with two free paramteters, $V_*$ and $\alpha_*$, to reproduce the observed cold gas fractions in spirals as a function of luminosity \citep{col94, col00}. 
The star formation timescale that does not depend on the dynamical timescale (i.e. redshift) has been supported from analysis on the formation of quasars \citep{kh00}, the evolution of damped Ly$\alpha$ systems \citep{spf01}, and the galaxy number counts \citep{nag01}.  
Since the value of $\alpha_*$ is usually assumed to be less than -1, the star formation timescale is longer for smaller galaxies.
This naturally explains the galaxy downsizing by keeping the gas in small galaxies until today, though the physical origin of the form of the star formation timescale is unclear.

\subsubsection{Merging of galaxies}

When two or more dark halos merge to form a new halo, the hot gas components in the progenitor halos are immediately merged and constitute a hot gas component in the new halo. In contrast, galaxies do not merge immediately. 
The central galaxy of the main progenitor halo is defined as the central galaxy of the new halo and the rest of galaxies are regarded as satellite galaxies. 
Satellite galaxies sink to the halo center owing to the dynamical friction and merge to the central galaxy. 
The satellite galaxies could merge each by random collision \citep{sp99}.
If the merging two galaxies have similar mass, the merger is considered as the {\it major merger} and a {\it starburst} occurs. 
The new galaxy becomes a pure bulge galaxy. In the case of a minor merger, the disk of the larger galaxy is not destroyed and a merging satellite is absorbed into either the disk or the bulge component of the larger galaxy. 
Note that some authors introduced small starbursts induced by minor mergers. For example \citet{on03} claimed that a starburst induced by a minor merger is necessary component in order to form the galaxy population with intermediate morphology. 

\subsubsection{Feedback}

Feedback processes arguably have the largest impact on the form of the theoretical predictions for galaxy properties, whilst at the same time being amongst the most difficult and controversial phenomena to model. 
The most common form of feedback used in models is the ejection of cold gas from a galactic disk by a supernova (SN) driven wind \citep[e.g.][]{lar74, ds86}.
The reheated cold gas could be blown out to the hot gas halo, from which it may subsequently recool (sometimes called {\it retention} feedback), or it may even be ejected from the halo ({\it ejection} feedback) and left unable to cool until it is incorporated into a more massive halo later. The ejection rate of the cold gas by SN driven wind is parameterized as follows:
\begin{equation}
\dot{M}_{\rm ej} = \left(\frac{V_{\rm c}}{V_{\rm hot}}\right)^{\alpha_{\rm hot}} \dot{M}_*,
\end{equation} 
where $V_{\rm hot}$ and $\alpha_o\rm hotp$ are free parameters. 
Considering the fact that the potential wells are shallower in smaller galaxies, the value of $\alpha_{\rm hot}$ should be negative. 
From energy balance, $\alpha_{\rm hot} = -2$ is obtained. 
However many authors used much stronger dependence on the circular velocity (i.e. $\alpha_{\rm hot} < -2$) in order to reproduce the faint end slope of luminosity functions \citep[e.g.][]{col94, ny04}. 

\citet{ben03} carried out a systematic study of the impact of various feedback mechanisms on the form of the predictions for the galaxy luminosity function.
The standard SN driven winds, while helping to reduce the number of faint galaxies to match observations, were found to overproduce bright galaxies. 
Stronger feedback of this mode would tend to weaken the break in the predicted luminosity function rather than enhance it by wiping out galaxies around $L_*$.
The thermal conduction and the {\it ejection} mode of feedback (superwinds) did the better job. 
However they had to assume an unphysically high conductivity in the conduction model and an implausibly high efficiency of feedback in the superwind model. 
\citet{cro06} and \citet{bow06} implemented simple AGN feedback schemes into their models, respectively. 
By assuming the {\it radio mode} feedback from AGN quenches gas cooling in quasi-hydrostatic hot halos, they reproduced the break in the luminosity function.

\subsubsection{Chemical evolution}

Chemical evolution of gas and stars in a galaxy is important because 
(i) the cooling rate of the gas, $\Lambda(T, Z)$, strongly depends on the gas metallicity \cite[e.g.][]{sd93}; 
(ii) the metallicity with which stars are born has an impact on the luminosity and colors of the stellar population; 
and (iii) the optical depth of a galaxy, which determines the extinction of starlight due to dust, scales with the mtallicity of its cold gas. 
As stars evolve, they return material to the interstellar medium (ISM) with an enhanced metallicity as the form of stellar winds or SN explosions. 
The amount of gas returned to the ISM per unit mass of stars forms is there for dependent on the form of the initial mass function (IMF) of the stellar population. 

The first attempt to follow the chemical evolution of galaxies in semi-analytic models considered type II SNe, using instantaneous recycling approximation. 
While the yield of metals is determined by the chosen IMF, the effective yield of metals depends on a number of differing metallicities such as (i) the mixing of hot gas reservoirs of differing metallicities by halo mergers, (ii) accretion of gas by cooling to the disc cold gas which could have different metallicity, and (iii) feedback processes which deliver metals in the cold gas to the hot halo gas.  

Recently type Ia SNe have been included in semi-analytic models. 
\citet{tho99} was the first to consider the delayed enrichment due to SNe Ia by using star formation histories extracted from the model of \citet{kwg93}, but neglecting any inflow or outflow of gas and metals. 
\citet{no06} produced the first semi-analytic model which self-consistently integrated the impact of SNe Ia, using the simplification of assuming a fixed time delay for SN Ia explosions.   
\citet{nag05a, nag05b} performed the first fully consistent calculation including both type II and Ia Sne in the semi-analytic models.

\subsection{Successes and failures}

Semi-analytic models now successfully reproduce the present day galaxy luminosity functions by invoking spurewinds \citep{ben03} or AGN feedback \citep{cro06, bow06}. 
\citeauthor{bow06} also reproduced luminosity functions at higher redshifts. 
However readers should bear in mind that any phenomenon which leads to a model successfully matching the observed break in the luminosity function uses up a dangerously high fraction of the energy released by supernova explosions. 
We also have to wait to see whether the models which invoke AGN feedback can account for the observed AGN activities. 

Semi-analytic models have a difficulty to match the zero-point of the Tully-Fisher relation and the normalization of the luminosity function at the same time. 
The difficulty lies on the fact that the CDM halos are too centrally concentrated and therefore it might be a genuine problem of the CDM. 
Scaling relations such as the Tully-Fisher relation rely on the calculation of the scale lengths of galaxies. 
Semi-analytic models assume that hot halo gas has the same specific angular momentum as its host halo and the angular momentum is conserved when gas becomes cold and accretes to a disk. 
However, numerical simulations have shown that angular momentum of gas can be transferred to the dark matter. 
Moreover, \citet{oka05} showed that the direction of the angular momentum vector of the main progenitor halo changed significantly during galaxy formation. 

\citet{bau05} were able to reproduce the number counts of sub-millimeter selected galaxies and the luminosity function of Lyman-break galaxies, whiles retaining a fair match to the present day optical and far infrared luminosity functions. 
However, this was only possible with the controversial assumption of a top-heavy IMF for star formation in merger driven starbursts. 
This model can also account for the metallicity of the intracluster medium and elliptical galaxies \citep{nag05a, nag05b}, though \citet{nag05b} failed to reproduce the observed trend of $\alpha/Fe$ increasing with the velocity dispersion of the galaxy. 
So far there is no explanation for this trend in the framework of the hierarchical galaxy formation. 

\section{Gasdynamic simulations}

Another powerful tool to study galaxy formation in cosmological context is gasdynamic simulations. 
In the gasdynamic simulations gas dynamics is solved based on governing equations. 
This is the major advantage of the gasdynamic simulations over the semi-analytic models in which gas distribution is {\it assumed}. 
There are two principle algorithm in common use to follow the hydrodynamics of gas in an expanding universe: particle-based Lagrangian schemes, which employ a technique called smoothed particle hydrodynamics (SPH) \citep[e.g.][]{mon92, chp95, pea99, sh03a}, and grid-based Eulerian schemes \citep[e.g.][]{ryu93, co99}. 
To date, most of gasdynamic simulations of galaxy formation have been peformed by SPH, because its Lagrangian nature is sutable to follow the collapse of gas clouds in cosmological volume.  

Prescriptions used in gasdynamic simulations in order to treat baryonic physics such as star formation and feedback are surprisingly similar to those used in semi-analytic models. The main difference is that such processes are formulated based on the local quantities in simulations rather than global quantities such as the circular velocity of the halo.
For example, the rate of change of the internal specific energy of $i$-th particle by cooling is calculated as
\begin{equation}
\dot{u}_i = -\frac{\Lambda(\rho_i, T_i, Z_i)}{\rho_i}, 
\end{equation}
where $u_i$, $\rho_i$, $T_i$, and $Z_i$ are the specific internal energy, density, temperature, and metallicity of the $i$-th particle, respectively.  
Note that in Eq.(\ref{COOLING}), the virial temperature of the halo and the mean metallicity of the hot halo gas are used.  
Although many simulations followed the chemical evolution, only a handful of studies \citep[e.g.][]{kg05, oka05} employed the metallicity dependent cooling function and other use the cooling rate of the primordial gas \citep[e.g.][]{aba03, som03, rob04, gov07} in spite of the fact that the cooling rate of the gas with the solar metallicity is more than an order of magnitude higher than that of the primordial gas. 
Usually, the inverse Compton cooling and effects of time evolving uniform ultraviolet background are also considered in the energy equation. 

Dense ($\rho > \rho_{\rm sf} \simeq 0.1$ cm$^{-3}$) and cold ($T \sim 10^4$ K) gas is eligible to form stars. 
Star formation rate of the $i$-th particle is given by
\begin{equation}
\dot{\rho}_* = C_* \frac{\rho_i}{t_{\rm dyn}} \propto \rho_i^{1.5}, 
\label{TSTAR}
\end{equation}
where $C_*$ is the dimensionless star formation efficiency parameter. 
To match the observed Kennicutt relation \citep{ken98}, $C_* \simeq 0.05$ is often used \citep[e.g.][]{ns00b}. 

Each star particle represents a single stellar population. 
Thus evolved stars explode as SNe.
Instantaneous recycling is often assumed. 
Some authors relaxed this approximation and calculated the number of SNe and returned mass as a function of the SSP's age \citep{oka05, oka07, gov07}. 
They also included SNe Ia. 
The feedback energy is distributed to the surrounding gas. 
It was found that depositing energy in the form of thermal energy had almost no effect because the energy was quickly radiated away in dense regions \citep{kat92}.
Note that since the cooling function depends on the temperature, not only the amount of energy given to the gas but also the total mass of gas to which the energy is given affect the thermal evolution. For example, giving all the feedback energy to the nearest one particle has much stronger effect than giving it to 50 particles.
To circumvent this problem, feedback energy is often given by the form of the kinetic energy \citep[e.g.][]{ns00b, sh03a}or cooling for the particles that receive the feedback energy is shutting-off for a while ($\sim 10$ Myr) \citep{tc01}. 

A self-consistent AGN feedback model was firstly incorporated into cosmological simulations of galaxy formation by \citep{oka07} in which they discriminated the quasar mode and radio mode based on the mass accretion rates onto the central blackholes. 
\citet{sij07} showed that such a discrimination successfully produced bright red galaxies.  

\subsection{Formation of spiral galaxies and effects of feedback}

\begin{figure}
\begin{center}
\includegraphics[width=11cm]{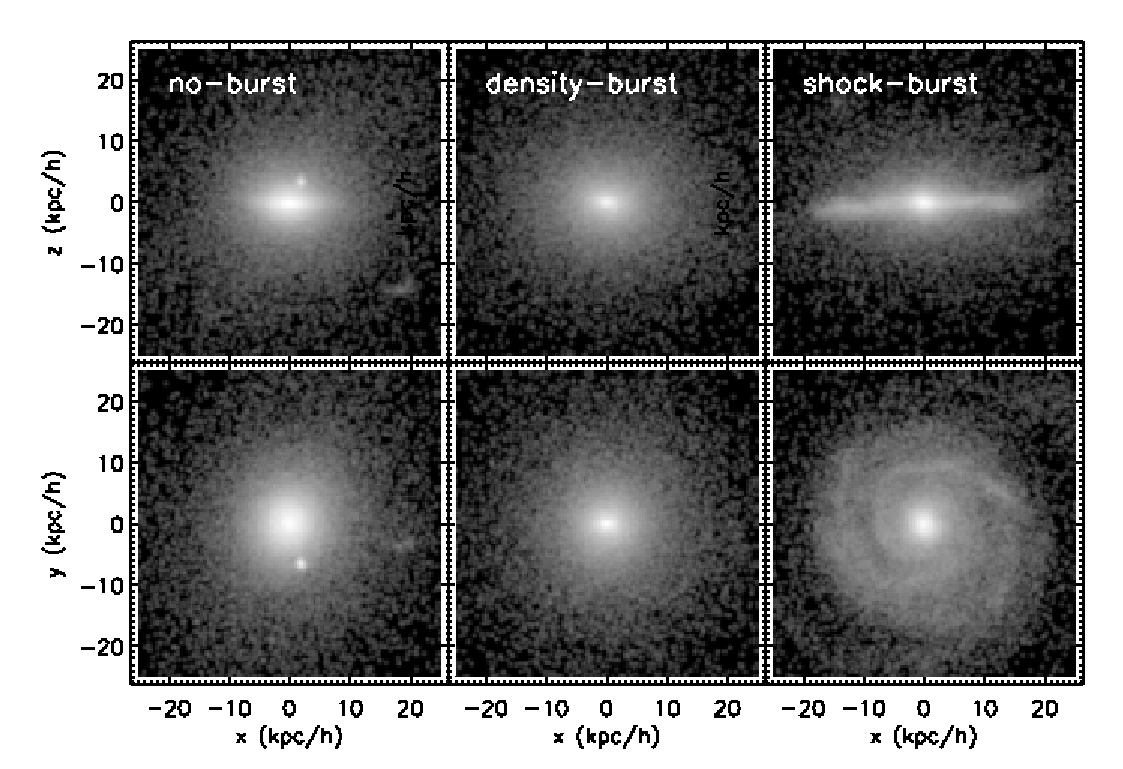}
\end{center}
\caption{Stellar distributions at $z = 0$. The no-burst, density-burst and shock-burst models are shown from left to right. 
  The edge-on and face-on views of stars are given in upper and lower panels, respectively. 
The brightness indicates the projected mass density, and the same scaling is used for each model. 
All of these galaxies are obtained from the same initial conditions.}
\label{STARS}
\end{figure}
The first attempts to simulate the formation of a spiral galaxy from CDM initial conditions generally failed, producing objects with overly centrally concentrated gas and stars. 
It became immediately apparent that the root cause of this problem was a net transfer of angular momentum from the baryons to the dark matter halo during the aggregation of the galaxy through mergers \citep{nb91, nw94, nfw95}. 
This is known as the {\t angular momentum problem}. 
It was suspected from the start that its solution was likely to involve feedback processes that would regulate the supply of gas to the galaxy.  

More recent simulations within the CDM framework have produced more promising disk galaxies. 
\citet{tc01} obtained a disk galaxy having reasonable size by assuming that gas heated by SN explosions can adiabatically expand without radiative loss of energy for a while. 
There simulation, however, stopped at $z = 0.5$. 
\citet{sn02} did not assume such strong feedback and found that a broad range of galaxy morphologies could be produced by gas accretion and galaxy mergers.
In related work, \citet{aba03} obtained a disk galaxy which resembled observed early-type disk galaxies. 
\citet{som03} were also able to generate a variety of morphological types, including disks, by assuming high star formation efficiency and strong feedback at high redshift to prevent early collapse of baryons. 
\citet{gov07} used the feedback model similar to \citet{tc01} in order to match the Tully-Fisher relation. 
There Tully-Fisher relation was, however, calculated from rotation speed of stars younger than 4 Gyr-old, which is much older populations than those observed by H$\alpha$. In fact, the rotation speed of the cold gas in their galaxy was too fast compared with observations.  
\citet{rob04} adopted a multiphase model for the ISM which stabilized gaseous disks against the Toomre instability, and produced a galaxy with an exponential surface brightness profile but insufficient angular momentum. 
These simulations, however, did not take the metallicity effects on cooling into account.

\citet{oka05} also assumed a multiphase model for the ISM and a top-heavy IMF for stars formed in starbursts, as required for semi-analytic models to match the number counts of bright submilimeter galaxies \citep{bau05} and metal contents in the intracluster medium and elliptical galaxies \citep{nag05a, nag05b}.  
By varying the criteria for starbursts in their simulations, \citet{oka05} ware able to produce galaxies with a variety of morphological types, from ellipticals to spirals, starting from exactly the same initial conditions (Fig.~\ref{STARS}). 

\subsubsection{The density profiles}
\begin{figure}
\begin{center}
\includegraphics[width=9cm]{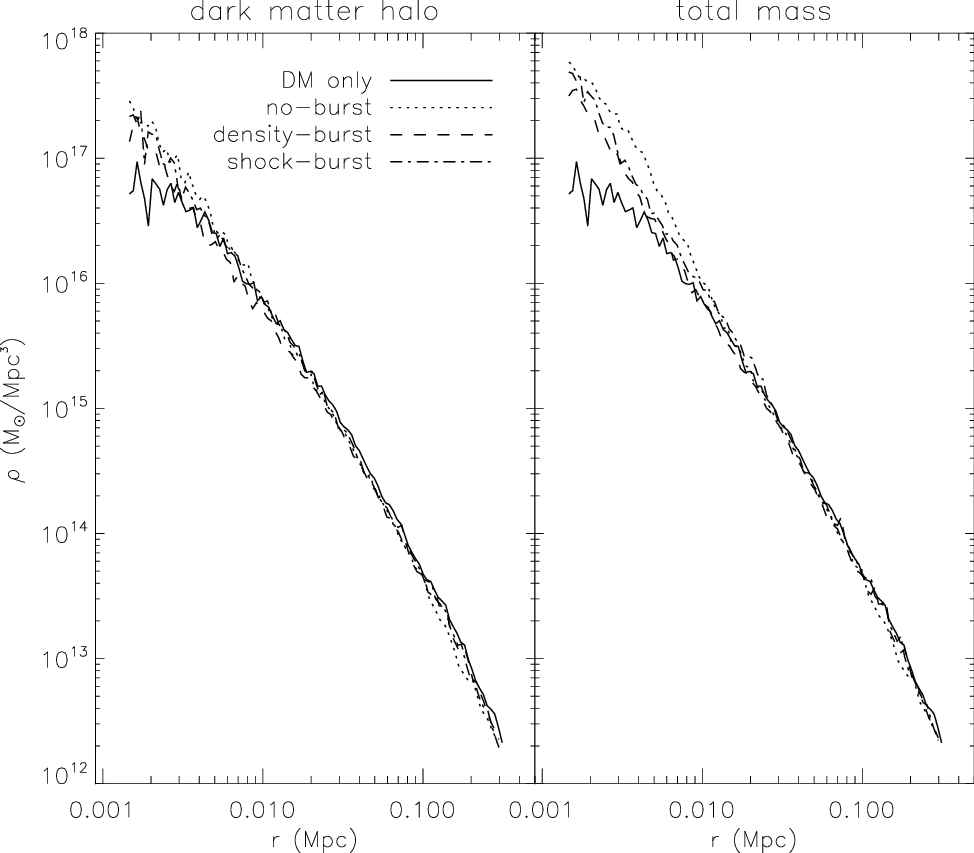}
\end{center}
\caption{
The left panel shows density profiles of dark halos for the DM only, {\it no-burst}, {\it density-burst}, and {\it shock-burst} simulations by the solid, dotted, dashed, and dot-dashed lines, respectively. 
The right panel is the same as the left one but for the total mass density profiles. 
}
\label{PROFILES}
\end{figure}
In Fig.~\ref{PROFILES}, density profiles of the host dark matter halos of the three galaxies in Fig.~\ref{STARS} are shown. The profile obtained by the simulation without baryonic component ($\Omega_{\rm b} = 0$) is also shown as the DM-only simulation. It is clear that the dissipatinal nature of the baryons makes the inner density profile steeper. The no-burst galaxy which has the weakest feedback effects has the steepest profile.

\subsubsection{The evolution of angular momentum}
\citet{zof07} analyzed two of the galaxies simulated by \citet{oka05}, the no-burst galaxy (left panel in Fig.~\ref{STARS}) and the shock-burst galaxy (righ). 
The no-burst galaxy is referred as the bulge-dominated galaxy and the shock-burst galaxy is called the disk-dominated galaxy. 
They traced the time evolution of the angular momentum of three {\it Lagrangian} components: the dark matter particles lie within the virial radius at $z = 0$ (halo), the 10\% most bound dark matter particles within the halo (inner halo), cold gas and stars within $0.1 r_{\rm vir}$ (galaxy). 
The tidal torque theory predicts that the amplitude of the angular momentum of a Lagrangian region evolves in proportion to $a^{3/2}$ until the region reaches the maximum expansion. Then the region starts to collapse and is decoupled from the tidal field, and thus the angular momentum stays constant \citep{whi84}.  
\begin{figure}[htb]
\begin{center}
\includegraphics[width=8cm]{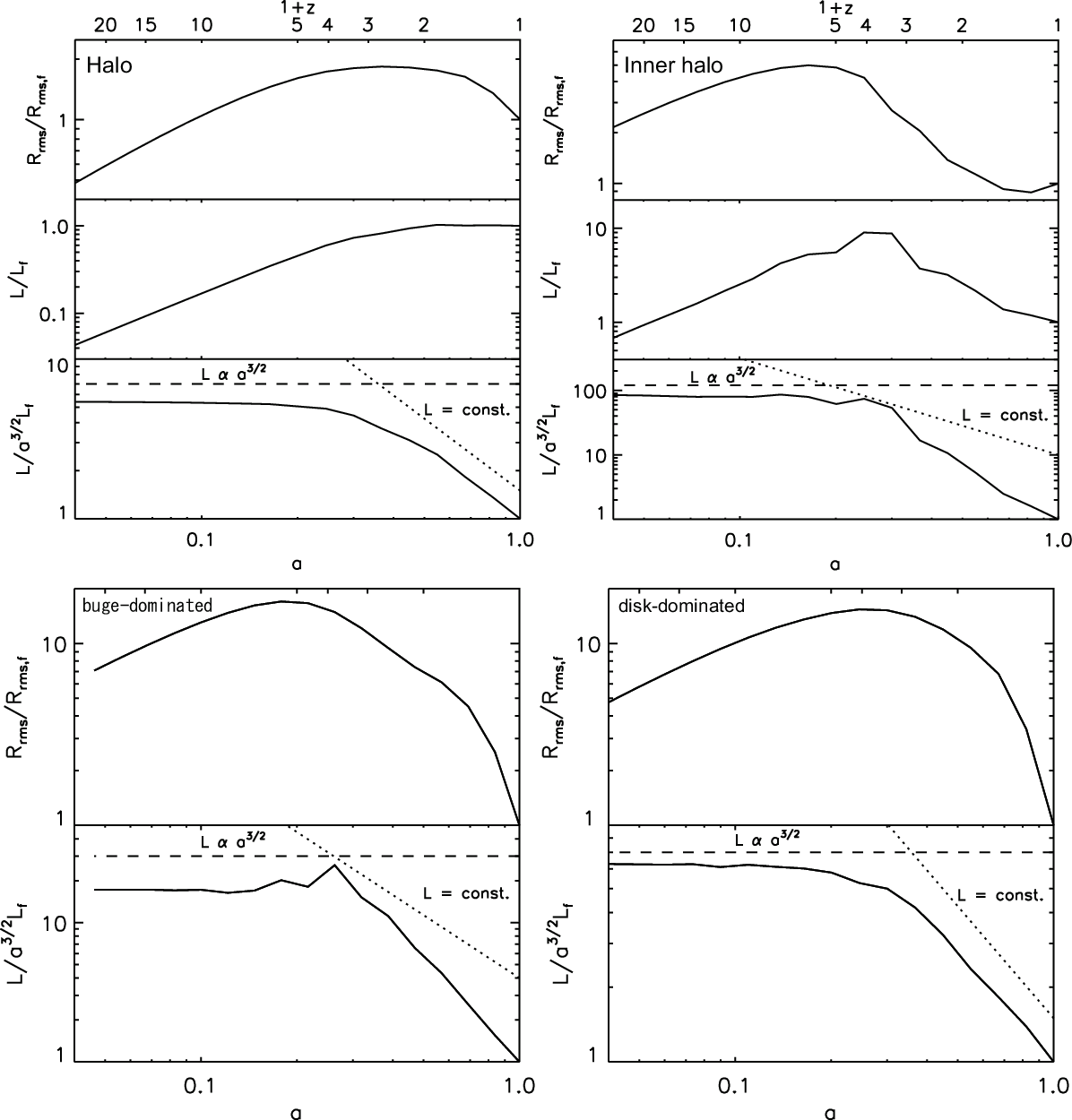}
\end{center}
\caption{
  Upper-left: the evolution of the {\it rms} radius (top panel) and the specific angular momentum (middle and bottom panels) of the dark mater particles that lie within $r_{\rm vir}$ at $z = 0$. In the bottom panel the dashed line indicated $L \propto a^{3/2}$ and the dotted line $L = const.$ 
    Upper-right: same as the upper-left but for the particles that makes the inner dark matter halo. 
    Lower-left: Evolution of the {\it rms} radius (upper panel) and specific angular momentum evolution (lower-panel) for the baryonic component of the bulge-dominated galaxy. Lower-right: same as the lower-left but for the disk-dominated galaxies.
    All physical quantities are normalized to their values at the present day.
}
\label{AM}
\end{figure}

The angular momentum evolution of the halo component (upper-left panel of Fig.~\ref{AM}) clearly shows two distinct phases, indicated by the dashed ($L \propto a^{3/2}$) and dotted ($L = const.$) lines. 
The inner halo shows quite different evolution (upper-left panel). Since this subregion has a higher overdensity than the halo as a whole, it reaches maximum expansion earlier. Until $z = 0$, the inner halo loses 90\% of the angular momentum by mergers. 
Lower panels in Fig.~\ref{AM} show angular momentum evolutions of galaxies (cold baryons) for the bulge-dominated galaxy (left) and disk-dominated galaxy (right). 
The evolution of the bulge-dominated galaxy (lower-left panel) is surprisingly similar to that of the inner halo (upper-right). This implies that baryons that are contained in the bulge-dominated galaxy are concentrated in the dense subsystems of dark matter at high redshift and lose their angular momentum as the inner halo loses its angular momentum during the collapse. 
On the other hand the evolution of the disk-dominated galaxy (lower-right) resembles that of the halo as a whole (upper-left). 
Strong feedback from the stellar population with a top-heavy IMF avoids early collapse of baryons and therefore baryons are diffusely distributed as the hot halo gas. 
When this gas cools, it accretes to the disk keeping its angular momentum. 
Thus, the suppression of the early collapse of baryons by some form of feedback is a key to solve the angular momentum problem. 
Interestingly, the model used in the disk-dominated galaxy (shock-burst simulation in \citet{oka05}) was also able to reproduce the luminosity function of the Milky Way satellites \citep{lib07}. 

\section{Conclusions}

I have presented results on recent efforts to model the formation and evolution of galaxies in hierarchically clustering universes using two complementary techniques: semi-analytic models and gasdynamic simulations. 
Feedback processes are the most important ingredient in these models which have reproduced the observed luminosity function and solved the angular momentum problem and the satellite problems. 
It is interesting that the feedback recipes that can solve the angular momentum problem also solve the satellite problem simultaneously \citep{lib07, gov07}. 
However in both techniques a dangerously high fraction of the energy released by SNe is used up. 
Sometimes more than 100\% of energy is used in simulations.
An excuse for doing so is that the ISM has a multiphase structure and therefore the cooling should not be as effective as in the simulations where the multiphase structure is not resolved. 
This uncertainty can be easily removed if we use the resolution that is sufficiently high to resolve star forming gas with density as high as $n_{\rm H} \sim 100$ cm$^3$ and temperature as low as $\sim 100$ K. 
Simulations of the ISM on the galactic scales have recently started \citep{tb06, rk07, sai07}. 
\citet{sai07} found that in these high resolution simulations star formation is not controlled by the star formation time-scale defined in equation (\ref{TSTAR}) but regulated by the time scale of gas supply form low density regions to high density regions, which is $\rho/\dot{\rho}  \simeq 5 \times t_{\rm dyn}$.
While there is still long way to go, resolving multiphase structure of the ISM in cosmological simulation will be a crucial step to make our {\it physically motivated} models to {\it models based on physics}. 
The radiation feedback has been largely ignored or only crudely included in our models. 
\citet{mqt05} suggested that large scale galactic winds seen around local and high redshift starburst galaxies are driven by radiation pressure rather than energy from SN explosions.  
To test this hypothesis, we have to carry out radiation hydrodynamics in galaxy formation simulations. 
Semi-analytic models also need refinements. 
While they have already have many parameters and are well-complicated, they are still too simple to take dynamical effects such as angular momentum transfer into account. 
Close interactions between simulations and semi-analytic models will be required for further improvements and better understanding of galaxy formation.

\acknowledgements %%% Text of acknowledgements runs on after this command.

The author thanks the organizers for inviting me to speak on this topic. 
He is indebted to Takayuki Saitoh for allowing me to show results in preparation. 
He also acknowledges support from PPARC. 
The simulations were carried out at the Cosmology Machine at the ICC, Durham.

%%% THE BIBLIOGRAPHY
%%%
%%% CONSULT SECTION 3 OF "INSTRUCTIONS FOR AUTHORS" FOR HOW TO USE NATBIB.
%%% AUTHORS ARE ENCOURAGED TO USE EITHER THE "THEBIBLIOGRAPY" ENVIRONMENT
%%% BY UNCOMMENTING (DELETING THE "%" SYMBOL) THE COMMANDS BELOW, OR BY
%%% USING THE BIBTEX ENVIRONMENT. TO FIND OUT WHICH IS APPLICABLE TO YOUR
%%% CONTRIBUTION, CONSULT THE VOLUME EDITORS FOR YOUR PROCEEDINGS.
%%%

% \bibliography{okamotoapj}

\end{document}